\documentclass[aps,pra,nofootinbib,twocolumn,notitlepage,superscriptaddress,showpacs,showkeys]{revtex4-2}
\usepackage{amsmath,amssymb,amsthm,mathrsfs,amsfonts,dsfont}
\usepackage{graphicx}
\usepackage{enumerate}
\usepackage{color}
\usepackage{braket}
\usepackage{todonotes}
\usepackage{lineno}
\usepackage{soul}
\usepackage{subfigure}
\graphicspath{{figures/}}

\usepackage[bookmarks=false,colorlinks,citecolor=blue,
linkcolor=blue,anchorcolor=blue,urlcolor=blue
]{hyperref}

\begin{document}

\title{Realigned Hardy's Paradox}
\author{Shuai Zhao}
\affiliation{School of Cyberspace, Hangzhou Dianzi University, Hangzhou 310018, China}
\affiliation{Hefei National Research Center for Physical Sciences at the Microscale and School of Physical Sciences, University of Science and Technology of China, Hefei, 230026, China}
\affiliation{CAS Center for Excellence in Quantum Information and Quantum Physics, University of Science and Technology of China, Hefei, 230026, China}
\author{Qing Zhou}
\author{Si-Ran Zhao}
\author{Xin-Yu Xu}
\author{Wen-Zhao Liu}
\affiliation{Hefei National Research Center for Physical Sciences at the Microscale and School of Physical Sciences, University of Science and Technology of China, Hefei, 230026, China}
\affiliation{CAS Center for Excellence in Quantum Information and Quantum Physics, University of Science and Technology of China, Hefei, 230026, China}
\author{Li Li}
\author{Nai-Le Liu}
\author{Qiang Zhang}
\affiliation{Hefei National Research Center for Physical Sciences at the Microscale and School of Physical Sciences, University of Science and Technology of China, Hefei, 230026, China}
\affiliation{CAS Center for Excellence in Quantum Information and Quantum Physics, University of Science and Technology of China, Hefei, 230026, China}
\affiliation{Hefei National Laboratory, University of Science and Technology of China, Hefei, 230026, China}
\author{Jing-Ling Chen}
\email{chenjl@nankai.edu.cn}
\affiliation{Theoretical Physics Division, Chern Institute of Mathematics, Nankai University, Tianjin 300071, China}
\author{Kai Chen}
\email{kaichen@ustc.edu.cn}
\affiliation{Hefei National Research Center for Physical Sciences at the Microscale and School of Physical Sciences, University of Science and Technology of China, Hefei, 230026, China}
\affiliation{CAS Center for Excellence in Quantum Information and Quantum Physics, University of Science and Technology of China, Hefei, 230026, China}
\affiliation{Hefei National Laboratory, University of Science and Technology of China, Hefei, 230026, China}

\begin{abstract}
Hardy's paradox provides an all-versus-nothing fashion to directly certify that quantum mechanics cannot be completely described by local realistic theory. However, when considering potential imperfections in experiments, like imperfect entanglement source and low detection efficiency, the original Hardy's paradox may induce a rather small Hardy violation and only be realized by expensive quantum systems. To overcome this problem, we propose a realigned Hardy's paradox. Compared with the original version of Hardy's paradox, the realigned Hardy's paradox can dramatically improve the Hardy violation. Then, we generalize the realigned Hardy's paradox to arbitrary even $n$ dichotomic measurements. For $n=2$ and $n=4$ cases, the realigned Hardy's paradox can achieve Hardy values $P(00|A_1B_1)$ approximate $0.4140$ and $0.7734$ respectively compared with $0.09$ of the original Hardy's paradox. Meanwhile, the structure of the realigned Hardy's paradox is simpler and more robust in the sense that there is only one Hardy condition rather than three conditions. One can anticipate that the realigned Hardy's paradox can tolerate more experimental imperfections and stimulate more fascinating quantum information applications.

\end{abstract}

\pacs{03.65.Ud, 03.67.HK, 03.67.-a}
\keywords{ Quantum theory, Bell nonlocality, Bell inequality, Hardy's paradox,}

\maketitle

\section{Introduction}
Bell nonlocality~\cite{Bell1964,Clauser1969,Clauser1974,Brunner2014}, which states that the predictions by quantum theory (QT) cannot be fully explained by the local-realistic theory~\cite{Einstein1935,Bohm1952A}, has deeply stimulated our cognitions on the physical world. Besides the fundamental interest, Bell nonlocality also lays the foundation for a variety of fascinating quantum information applications. Indeed, in a model that reveals Bell nonlocality, the measurement results cannot be fully predicted by pre-sharing local variables. Therefore, once Bell nonlocality is observed in a quantum information processing task, the measurement outputs cannot be fully determined by a malicious third party no matter what mechanism underlying the devices, i.e., device-independent security~\cite{Acin2007Device}. For instances, Bell nonlocality guarantees the security of device-independent quantum key distribution~\cite{Acin2007Device,Lim2013Device,Tan2020Advantage,Cao2016Performance}, device-independent quantum random number generation~\cite{pironio2010random,Acin2012Randomness,Ramanathan2016Randomness} and self-testing of quantum systems~\cite{mayers1998quantum,mayers2004self}.

Practically, Bell inequalities are widely adopted to demonstrate Bell nonlocality and recently the related quantum information applications have also been pushed into the regime of experimental realization~\cite{Li2010Gisin,Brunner2014,liu2018device,liu2021device,Li2021Experimental,Liu2022Toward,zhang2022device,nadlinger2022experimental}. Besides Bell inequalities, other effective ways to certify Bell nonlocality known as Bell's theorem without inequality are also well studied, such as Greenberger-Horne-Zeilinger paradox~\cite{Greenberger1990Bell}, Hardy's paradox~\cite{Hardy1993Nonlocality}, etc. Here, we focus on the Hardy's paradox, which proves an all-versus-nothing fashion to demonstrate the Bell nonlocality and is well known as the simplest version of Bell's theorem~\cite{Mermin1994Quantum}. Briefly, when the Hardy's conditions are satisfied, the Hardy value happens to be zero for any local realistic theories, while, it will achieve foremostly a value of $(5\sqrt{5}-11)/2$ for the quantum theory which is a paradox that implies a violation against the local realistic theory. Based on the Hardy's paradox, several important quantum information applications have been proposed such as quantum randomness generation and device-independent quantum key distribution~\cite{Li2015Device,Rahaman2015Quantum,Rahaman2015Device}.  Experimentally, there are potential loopholes in Bell tests~\cite{Brunner2014,Zhao2019Higher} including the Hardy's paradox test, such as the well discussed \emph{locality loophole}~\cite{Brunner2014}, \emph{freedom-of-choice loophole}~\cite{Scheidl19708Violation} and \emph{detection loophole}~\cite{Pearle1970Hidden,Brunner2007Detection}, which may result in invalid conclusion of Bell tests as well as the related quantum information applications\cite{Acin2007Device,Lim2013Device,Tan2020Advantage,Acin2012Randomness,Ramanathan2016Randomness}. Intuitively, when closing the detection loophole, the violation of the Hardy's paradox goes down as the detection efficiency $\eta$ decreases. The lower bound for demonstrating the Bell nonlocality with the Hardy's paradox is presented to be $\eta=2/3$, which is the same as that using the Clauser-Horne-Shimony-Holt (CHSH) inequality~\cite{Clauser1969,Brunner2014,eberhard1993background,hwang1996detection}. Up to now, the loophole-free demonstration of the Hardy's paradox remains difficult due to the high requirement of experiment systems. To overcome this problem, one may use pure enough entanglement sources and detectors with high enough efficiency, which are still expensive even for proof-of-principle demonstrations. Another way is to develop new Hardy paradox scenarios which can tolerant more imperfections. In this work, we propose the realigned Hardy's paradox which can dramatically improve the Hardy value compared with the original Hardy's paradox. Meanwhile, the Hardy conditions are simplified to be only one condition in the realigned Hardy's paradox which is more robust. To reveal more fascinating aspects of Bell nonlocality, we generalize the realigned Hardy's paradox to be that with arbitrary $n$ dichotomic measurements($n$ is an even number) for Alice and Bob. As a demonstration, the violation of the realigned Hardy's paradox with $n=2$ and $n=4$ binary measurements are presented to be $0.4140$ and $0.7734$, respectively. Therefore, we can anticipate that the realigned Hardy's paradox can tolerate more experimental imperfections and stimulate more fascinating quantum information applications.

\begin{figure}[htbp]
\includegraphics[width =0.4\textwidth]{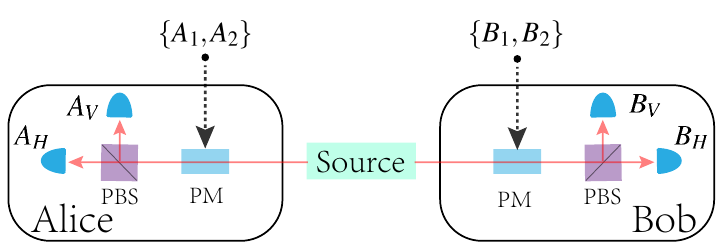}
\caption{The optical schematic setup for demonstrating Bell nonlocality using entangled pairs from an entangled source. $\{A_1, A_2\}$ and $\{B_1, B_2\}$ are the measurement settings for Alice and Bob. $A_H$, $A_V$, $B_H$ and $B_V$ label the corresponding detectors. BS: non-polarization beam splitter; HWP: half-wave plate; PM: polarization modulator; PBS: polarization beam splitter. }\label{fig_setup}
\end{figure}

\section{Hardy's Paradox}\label{Hardy_paradox_section}
The schematic setup in Fig.~\ref{fig_setup} is a typical optical setup for experimentally demonstrating Bell nonlocality.
There are an entangled source and two separated players (i.e., Alice and Bob) with two binary measurements~\cite{Brunner2014}.  Alice (Bob) chooses measurement settings $x\in\{A_1, A_2\}$ ($y\in\{B_1, B_2\}$) to make measurements on the particles within her (his) hand. The measurement outputs for Alice and Bob are binary, i.e., $i,j\in\{0,1\}$, respectively.
In terms of the local hidden variable (LHV) model, the conditional probabilities for the outcomes $i$ and $j$ conditioned on inputs $x$ and $y$ can be expressed as
\begin{equation}\label{LHVprob}
  P(ij|xy)=\int d \lambda q(\lambda)P(i|x,\lambda)P(j|y,\lambda),
\end{equation}
where $\lambda$ is the local hidden variable introduced by a local realistic theory, and $q(\lambda)$ is the probability distribution with $\int q(\lambda)d \lambda =1$. This probability distribution describes the behavior of local realistic theory, the set of which is a convex polytope~\cite{Brunner2014,goh2018geometry}. The original Hardy's paradox is expressed as
\begin{equation}\label{hardy_paradox_original}
  \begin{split}
    &P(00|A_2B_2)=0,\\
    &P(01|A_1B_2)=0,\\
    &P(10|A_2B_1)=0,\\
    &\rule{4cm}{0.05em}\\
    &P(00|A_1B_1)\overset{\text{LHV}}{=}0 ~(\overset{\text{QT}}{>}0),
  \end{split}
\end{equation}
where the first three equations above the line are termed as Hardy conditions, the probability below the line is Hardy value~\cite{Hardy1993Nonlocality}. For the LHV model, when the Hardy conditions are satisfied, one can prove that the Hardy value should always be $0$ by travelling over extreme points of the aforementioned convex polytope. While, for quantum theory, the Hardy value can happen to be maximally   $$P(00|A_1B_1)^{\text{max,QT}}=\frac{5\sqrt{5}-11}{2}\approx 0.09,$$
which shows violation for the local realistic theory~\cite{Hardy1993Nonlocality,Chen2013Hardy}.

Similar with Bell inequality tests,  there will also be aforementioned potential loopholes in the Hardy's paradox tests as well as the related quantum information applications\cite{Acin2007Device,Lim2013Device,Tan2020Advantage,Acin2012Randomness,Ramanathan2016Randomness,Brunner2014,Zhao2019Higher}. When closing all the potential loopholes, i.e., loophole-free Hardy's paradox test, it will result in a rather small violation which is not easy to be observed in experiment. To solve this problem, we propose a realigned Hardy's paradox which can dramatically improve the Hardy value for quantum theory. Next, we present the details about the realigned Hardy's paradox.

\section{Realigned Hardy's Paradox}
The original Hardy's paradox in Eq.~(\ref{hardy_paradox_original}) pioneered an all-versus-nothing way for demonstrating Bell nonlocality. While, as discussed above, there may be some difficulties when considering loophole-free Hardy's paradox test. Here, we present the realigned Hardy's paradox as a solution for the problem. To improve the violation of quantum theory, we propose the realigned Hardy's paradox,
\begin{equation}\label{realigned_Hardy_Paradox}
  \begin{split}
    &P(11|A_1B_1)+P(10|A_2B_2)+P(00|A_1B_2)\\
    &+P(11|A_2B_1)+P(11|A_1B_2)+P(00|A_2B_1)\\
    &+P(01|A_2B_2)=3,\\
   &\rule{7cm}{0.05em}\\
    &P(00|A_1B_1)\overset{\text{LHV}}{=}0~(\overset{\text{QM}}{>}0).
  \end{split}
\end{equation}
Intuitively,  there is only one Hardy condition compared with the original Hardy's paradox. When the Hardy condition is satisfied, the Hardy value must be $P(00|A_1B_1)=0$ for any local realistic theories. While, one can calculate the maximal Hardy value for quantum theory to be 
\begin{equation}
  P(00|A_1B_1)^{\text{max,QT}}\approx 0.4140.
\end{equation}
The proof is summarized in Methods (see section \ref{proof_Hardy_paradox_CHSH} for details). Recalling the Hardy value 0.09 of the original Hardy's paradox, it can dramatically improve the violation of the Hardy's paradox. Compared with the original Hardy's paradox, the realigned Hardy's paradox has following advantages. Firstly, the realigned Hardy's paradox can dramatically improve the Hardy value for the quantum theory, which has greater violation for the local realistic theory. Secondly, the three Hardy's conditions are casted to be only one condition, which can simplify the experiments to a great extent. Next, we present how to generalize the realigned Hardy's paradox to arbitrary even $n$ dichotomic measurements which will lead to much higher violations against local realistic theories compared with the original Hardy's paradox.

\section{Generalization to realigned Hardy's Paradoxes with more dichotomic measurements}
To reveal more fascinating aspects of Bell nonlocality and promoting the relevant quantum information applications, one important clue is to develop Bell tests with more measurement settings and more measurement outcomes~\cite{GISIN1999Bell}. Based on the schematic setup in Fig.~\ref{fig_setup}, we show that the above realigned Hardy's paradox can also be generalized to that with arbitrary even $n$ measurements for Alice and Bob.

In this part, we start from the well discussed two-qubit $n$-setting Abner Shimony (AS) inequality ($n$ is even number)~\cite{gisin2009bell,Collins_2004A,Avis_2006On,Yang2019Stronger} to construct the realigned Hardy's paradoxes with $n$ dichotomic measurements. The $\text{AS}$ inequality in its probabilistic form is formulated as
\begin{equation}\label{Inn22}
  \begin{aligned}
    I_{nn22}&=\sum_{i=1}^n\sum_{j=1}^{n-i+1}P(A_i=B_j)\\
    &+\sum_{i=2}^{n/2}(i-1)[P(A_i\neq B_{n-i+2})+P(A_{n+2-i}\neq B_i)]\\
    &+\frac{n}{2}P(A_{(n/2)+1}\neq B_{(n/2)+1})\\
    &\overset{\text{LHV}}{\leq}\frac{n^2+n}{2}\overset{\text{QM}}{\leq}\frac{\frac{(n+1)\sqrt{n(n+2)}}{3}+\frac{3n^2+2n}{4}}{2},
  \end{aligned}
\end{equation}
where $n$ is the number of measurement settings, $A_i\in\{A_1, A_2 \cdots A_n\}$ and $B_j\in\{B_1, B_2 \cdots B_n\}$ are measurement inputs in Alice's and Bob's sides respectively, $A_i=B_j$($A_i\neq B_j$) means that the measurement outcomes for measurement inputs $A_i$ and $B_j$ are the same (different). When $n=2$, the $\text{AS}$ inequality reduces into the CHSH inequality (see Eq.~(\ref{CHSH_proba})).

From the $I_{nn22}$ inequality of Eq.~(\ref{Inn22}), the probability terms are realigned to be Hardy value and Hardy condition to construct new realigned Hardy's paradox of the form
\begin{equation}\label{hardy_Inn22}
  \begin{split}
    &I_{nn22}-P(00|A_1B_1)=\frac{n^2+n}{2},\\
   &\rule{5cm}{0.05em}\\
    &P(00|A_1B_1)\overset{\text{LHV}}{=}0~(\overset{\text{QM}}{>}0).
  \end{split}
\end{equation}
Similar with the proof for $n=2$ case, the classical bound of the Hardy value $P(00|A_1B_1)=0$ can be directly obtained by analysis on the classical bound of the AS inequality. The upper bound for quantum theory is of more interests in our case. As an example, we take $n=4$ to show how to construct the realigned Hardy's paradox based on $I_{4422}$ inequalities. In the $I_{4422}$ inequality, the upper bound for LHV model is $10$. To construct the realigned Hardy's paradox based on the $I_{4422}$ inequality, we choose one of the probability terms as the Hardy value and the remaining probability terms as the Hardy condition. Specifically, we set $P(00|A_1B_1)$ to be the Hardy value and other probabilities of the $I_{4422}$ inequality as the Hardy condition. The realigned Hardy's paradox based on the $I_{4422}$ inequality is formulated as
\begin{equation}\label{Realigned_Hardy_I4422}
  \begin{split}
   &P(00|A_1B_2)+P(11|A_2B_3)+P(11|A_1B_3)\\
   &+2P(10|A_3B_3)+P(11|A_2B_2)+P(00|A_3B_1)\\
   &+P(10|A_2B_4)+P(00|A_4B_1)+P(11|A_3B_2)\\
   &+2P(01|A_3B_3)+P(00|A_1B_4)+P(11|A_1B_4)\\
   &+P(11|A_2B_1)+P(00|A_2B_1)+P(00|A_2B_3)\\
   &+P(11|A_3B_1)+P(00|A_2B_2)+P(01|A_2B_4)\\
   &+P(11|A_4B_1)+P(11|A_1B_1)+P(11|A_1B_2)\\
   &+P(00|A_1B_3)+P(00|A_3B_2)+P(01|A_4B_2)\\
   &+P(10|A_4B_2)=10,\\
    &\rule[-10pt]{7.5cm}{0.05em}\\
    &P(00|A_1B_1)\overset{\text{LHV}}{=}0~(\overset{\text{QM}}{>}0).
  \end{split}
\end{equation}
The right-hand side of Hardy condition is $10$ which is the upper bound of the $I_{4422}$ inequality. From the $I_{4422}$ inequality , once the Hardy condition is satisfied, the Hardy value must be $P(00|A_1B_1)=0$ for any local realistic theory. While, quantum theory can achieve a non-zero Hardy value. With the global search method, the maximal value for qubit system is
\begin{equation}
  P(00|A_1B_1)\approx 0.7734.
\end{equation}
It should be noted that the upper bound for quantum theory of this realigned Hardy's paradox may not be achieved by qubit system due to the Jordan's Lemma, which simplifies the problems involves Bell scenarios with $n$ parties $2$ dichotomic measurements to qubit systems~\cite{goh2018geometry}. Therefore, we use the NPA hierarchy method to bound the Hardy value for quantum theory from the outer side, which suggests $P(00|A_1B_1)\approx 0.7804$ (at the third level of the NPA method)~\cite{Navascu2008convergent}. Thus, the true quantum upper bound for $P(00|A_1B_1)$ should be a value in $[0.7734, 0.7804]$. The proof of the realigned Hardy's paradox based on the $I_{4422}$ inequality is similar with that of Eq.~(\ref{realigned_Hardy_Paradox}), and we present the details in Methods (see section \ref{proof_Hardy_paradox_CHSH}).

Compared with the original Hardy's paradox, it has improved the Hardy value from 0.09 to 0.7734 for qubit systems and has the simple structure of only one Hardy condition. Meanwhile, the realigned Hardy's paradox with arbitrary even $n$ measurements for Alice and Bob can be constructed. One can anticipate that the realigned Hardy's paradox with more measurements can tolerant more imperfections in experiments and have more fascinating quantum information applications such as device-independent quantum random number generation, quantum randomness amplification, etc.

\section{Discussion and Conclusion}
Hardy's paradox presents an essentially effective method to verify Bell nonlocality and lays an important foundation for quantum information applications. However, when considering potential experimental imperfections, the violation of the original Hardy's paradox may induce rather small violation and can only be observed by detectors with very high detection efficiency. To overcome this problem, we propose the realigned Hardy's paradox.  Firstly, the realigned Hardy's paradox can dramatically improve the Hardy value for the quantum theory, which has greater violation against the local realistic theory. Secondly, the three Hardy's conditions is casted to be only one condition, which can simplify the experiments to a great extent. Thirdly, based on the AS inequality,  the realigned Hardy's paradox can be generalized to be with arbitrary even $n$ dichotomic measurements for Alice and Bob, which reveals more fascinating aspects of Bell nonlocality and may facilitate the potential quantum information applications.  Meanwhile, this work provides an insightful method to convert Bell inequalities with arbitrary number of inputs and outputs to be a generalized Hardy's paradox. About the potential experimental performance under imperfect devices, it still needs further investigation. Due to larger violation of realigned Hardy's paradox, we anticipate it can tolerant more imperfections in experiment and leave it for future works.


\begin{table*}[htbp]
\caption{Optimized quantum states and measurement observables obtained from qubit systems for $n=2$ and $n=4$ cases. $\theta\in[-\pi,\pi]$ is the parameter for two-qubit entanglement state. $\alpha_1, \alpha_2, \alpha_3, \alpha_4\in[-\pi,\pi]$ and $\beta_1, \beta_2, \beta_3, \beta_4\in[-\pi,\pi]$ are angles for Alice and Bob's observables, respectively.}\label{tbl1}
\begin{tabular*}{0.7\textwidth}{c c c c c c c c c c c }
\hline
$n$ & $\theta$ & $\alpha_1$ & $\alpha_2$ & $\alpha_3$ & $\alpha_4$ & $\beta_1$ & $\beta_2$ & $\beta_3$ &$\beta_4$ & $P(00|A_1B_1)$\\
\hline
2 & 0.7968 & -0.1996 & 0.5901 &  &  & 0.1996 & -0.5901 &  &  & 0.4140\\
4 & 1.0793 & -1.5309 & 1.3084 & 2.1179 & 0.9181 & -1.6107 & -1.3084 & -2.1179 & -0.9181 & 0.7734\\
\hline
\end{tabular*}
\end{table*}

\section{Methods: Proof of the realigned Hardy's paradoxes}\label{proof_Hardy_paradox_CHSH}
\begin{sloppypar}
To prove the upper bound for local realistic theory, one can directly travel over all the extreme points of the polytope for LHV model (Sixteen extreme points in total for $n=2$ inputs cases). In this section, to be more intuitive, we present the proof of the realigned Hardy's paradox with the help of CHSH inequality as follows:
\begin{proof}
The CHSH inequality in its probability form is formulated as~\cite{Cabello2013Simple,Yang2019Stronger}
\begin{equation}\label{CHSH_proba}
\begin{split}
   &P(11|A_1B_1)+P(10|A_2B_2)+\\
   &P(00|A_1B_2)+P(11|A_2B_1)+\\
   &P(11|A_1B_2)+P(00|A_2B_1)+\\
   &P(01|A_2B_2)+P(00|A_1B_1)\\
   &\overset{\text{LHV}}{\leq} 3\overset{\text{QM}}{\leq}2+\sqrt{2}.
\end{split}
\end{equation}
In the CHSH inequality, the upper bound for LHV model is $3$. In the realigned Hardy's paradox, the Hardy value $P(00|A_1B_1)$ is the last probability of the CHSH inequality, and other probabilities of the CHSH inequality constitute the Hardy condition. From the CHSH inequality, once the Hardy condition fulfills the classical bound, the Hardy value must be $P(00|A_1B_1)=0$ for any local realistic theory. While, in quantum theory, one can achieve non-zero Hardy values. Specifically, the maximal Hardy value for quantum theory is $\displaystyle P^{max}_{\text{Hardy}}\approx 0.4140$, which can be calculated by numerical methods (See Table~\ref{tbl1}). Definitely, this value is suggested by both the global search method for qubit system and the NPA hierarchy method~\cite{Navascu2008convergent}.
\end{proof}
Similarly, one can prove the validation of the realigned Hardy's paradox with four measurement settings based on the $I_{4422}$ inequality.
When taking $n=4$, the AS inequality $I_{nn22}$ in main text is formulated as
\begin{equation}\label{I4422}
  \begin{aligned}
    I_{4422}&=P(00|A_1B_2)+P(11|A_2B_3)+P(11|A_1B_3)\\
    &+2P(10|A_3B_3)+P(11|A_2B_2)+P(00|A_3B_1)\\
    &+P(10|A_2B_4)+P(00|A_4B_1)+P(11|A_3B_2)\\
    &+2P(01|A_3B_3)+P(00|A_1B_4)+P(11|A_1B_4)\\
    &+P(11|A_2B_1)+P(00|A_2B_1)+P(00|A_2B_3)\\
    &+P(11|A_3B_1)+P(00|A_2B_2)+P(01|A_2B_4)\\
    &+P(11|A_4B_1)+P(11|A_1B_1)+P(11|A_1B_2)\\
    &+P(00|A_1B_3)+P(00|A_3B_2)+P(01|A_4B_2)\\
    &+P(10|A_4B_2)+P(00|A_1B_1)\\
    &\overset{\text{LHV}}{\leq}10\overset{\text{QM}}{\leq}7+\frac{5\sqrt{6}}{3}.
  \end{aligned}
\end{equation}
Then, we take one of the probability in Eq.~(\ref{I4422}) as the Hardy value, and other probabilities constitute Hardy condition. Meanwhile, the Hardy's condition is set to be equal to the upper bound for LHV model of the $I_{4422}$ inequality. Thus, the realigned Hardy's paradox with four measurement settings is of the form of Eq.~(\ref{Realigned_Hardy_I4422})
\begin{equation*}
  \begin{split}
   &I_{4422}-P(00|A_1B_1)=10,\\
   &\rule[-10pt]{5cm}{0.05em}\\
   &P(00|A_1B_1)\overset{\text{LHV}}{=}0~(\overset{\text{QM}}{>}0).
  \end{split}
\end{equation*}
Without causing confusion, we directly denote the present realigned Hardy's paradox with four measurement settings as the realigned Hardy's paradox based on the $I_{4422}$ inequality.
\begin{proof}
  From the $I_{4422}$ inequality of Eq.~(\ref{I4422}), the upper bound for any local realistic theory is $10$. The proof of the classical bound of this realigned Hardy's paradox is directly. While, for quantum theory, we obtain the maximal Hardy value for qubit system $P(00|A_1B_1)\approx 0.7734$ with numerical methods (See Table~\ref{tbl1}). Need to note that the upper bound for this realigned Hardy's paradox may not be achieved by qubit system due to the Jordan's Lemma~\cite{goh2018geometry}. Therefore, we use the NPA hierarchy method to bound the Hardy value for quantum theory from the outer side, which suggests $P(00|A_1B_1)\approx 0.7804$ (at the third level of the NPA method).
\end{proof}

To quantify the upper bound of the Hardy value $P(00|A_1B_1)$ for a general realigned Hardy's paradox of Eq.~(\ref{hardy_Inn22}), the NPA hierarchy method can be adopted to give an effective bound from the outside~\cite{Navascu2008convergent}. The problem to get the upper bound for the realigned Hardy's paradox is formulated as
\begin{equation}
  \begin{split}
    \text{max} \; &P(00|A_1B_1)\\
     s.t.\;~
           & I_{nn22}-P(00|A_1B_1)=\frac{n^2+n}{2},\\
           & \{P(ij|xy)\} \in \mathcal{Q}_l,
  \end{split}
\end{equation}
where $\{P(ij|xy)\}$ is the probability distribution with $i,j\in\{0,1\}, x\in\{A_1,A_2,\cdots, A_n\}$ and $y\in\{B_1,B_2,\cdots, B_n\}$. $\mathcal{Q}_l$ is the semi-definite programming relaxation of the set of quantum correlations at the $l$-th level of the NPA hierarchy method.

Here we present the model for qubit system. Without loss of generality, the two-qubit quantum state is of the form
\begin{equation}
  \psi(\theta)=\cos\theta|00\rangle+\sin\theta|11\rangle,
\end{equation}
where $\theta\in[-\pi,\pi]$ is adjustable parameter. We take the measurement observables in $X-Z$ plane that
 \begin{equation}
  \begin{split}
        A_{k_a}&=\left(
          \begin{array}{cc}
            \cos2\alpha_{k_a} & \sin2\alpha_{k_a} \\
            \sin2\alpha_{k_a} & -\cos2\alpha_{k_a} \\
          \end{array}
        \right), \\
        B_{k_b}&=\left(
          \begin{array}{cc}
            \cos2\beta_{k_b} & \sin2\beta_{k_b} \\
            \sin2\beta_{k_b} & -\cos2\beta_{k_b} \\
          \end{array}
        \right),
  \end{split}
  \end{equation}
where $k_a$ and $k_b$ are subscripts for measurement setting $\{A_1,A_2,\cdots, A_n\}$ and $\{B_1,B_2,\cdots, B_n\}$, respectively. $\alpha_{k_a}\in[-\pi,\pi]$ and $\beta_{k_b}\in[-\pi,\pi]$ are angles for corresponding measurement observables. When the observable for measurements are taken to be $x=A_{k_a}$ and $y=B_{k_b}$, the probability distribution for qubit system is
\begin{equation}\label{qubit_pro}
  P(ij|A_{k_a}B_{k_b})=Tr[\frac{\mathbf{I}+(-1)^iA_{k_a}}{2}\otimes\frac{\mathbf{I}+(-1)^jB_{k_b}}{2}\rho],
\end{equation}
where $\rho=|\psi(\theta)\rangle\langle\psi(\theta)|$ is the density matrix for Alice and Bob's qubit system, and $\mathbf{I}$ is the identity matrix. Thus, the problem to get the upper bound for the realigned Hardy's paradox using qubit system is formulated as
\begin{equation}
  \begin{split}
    \text{max} \; &P(00|A_1B_1)\\
     s.t.\;~
           & I_{nn22}-P(00|A_1B_1)=\frac{n^2+n}{2},\\
  \end{split}
\end{equation}
where the probability distribution $\{P(ij|xy)\}$ admits Eq.~(\ref{qubit_pro}).
Specifically, the optimized parameters are shown in Table~\ref{tbl1}.

\end{sloppypar}

\section*{Acknowledgments and Funding}
\begin{sloppypar}
We acknowledge Ravishankar Ramanathan for insightful discussion. This work has been supported by the National Natural Science Foundation of China (Grants No.62031024, No.11874346, No.12174375, No.12275136, No.11875167 and No. T2125010), the National Key R$\&$D Program of China (Grant No.2019YFA0308700), the Anhui Initiative in Quantum Information Technologies (Grant No.AHY060200) as well as the Innovation Program for Quantum Science and Technology (Grant No. 2021ZD0301100).
\end{sloppypar}

\section*{Conflict of interest}
The authors declare that they have no conflict of interest.


\nolinenumbers


%

\end{document}